\documentclass[a4paper]{jpconf}
\usepackage{graphicx}
\usepackage{latexsym}
\usepackage{bm}

\newcommand{\CGf}{\ensuremath{\mathcal{C}}}
\newcommand{\PGf}{\ensuremath{\mathcal{P}}}

\newcommand{\ave}[1]{\langle #1 \rangle}

\begin{document}
\title{Honeycomb lattice polygons and walks as a test of series analysis techniques}
\author{Iwan Jensen}
\address{ARC Centre of Excellence for Mathematics and Statistics of Complex Systems, 
Department of Mathematics and Statistics, 
The University of Melbourne, Victoria 3010, Australia}

\ead{I.Jensen@ms.unimelb.edu.au}

\begin{abstract}
We have calculated long series expansions for self-avoiding walks and polygons 
on the honeycomb lattice, including series for metric properties such as
mean-squared radius of gyration as well as series for moments of the 
area-distribution for polygons. Analysis of the series yields accurate estimates 
for the connective constant, critical exponents and amplitudes of honeycomb 
self-avoiding walks and polygons. 
The results from the numerical analysis agree to a high degree  of accuracy 
with theoretical predictions for these quantities.
\end{abstract}

\section{Introduction}

Self-avoiding walks (SAWs) and polygons (SAPs) on regular lattices are among 
the most important and classic combinatorial problems in statistical mechanics. 
SAWs are often considered in the context of lattice models of polymers while SAPs
are used to model vesicles. The fundamental problem is the calculation (up to translation) 
of the number of SAWs, $c_n$, with $n$ steps (SAPs, $p_n$, of perimeter $n$). As for many
interesting combinatorial problems, SAWs have exponential growth, $c_n \sim A\mu^n n^{\gamma-1}$, 
where $\mu$ is the connective constant, $\gamma$ is a critical exponent, and $A$ is a critical 
amplitude. A major challenge (short of an exact solution) is the calculation, or at least 
accurate estimation of, $\mu$, critical exponents and amplitudes. Here our focus is on
the numerical estimation of such quantities from exact enumeration data.  

The success of series expansions as a numerical technique has relied crucially on 
several of Tony Guttmann's contributions to the field a asymptotic series analysis.
In pioneering the method of differential approximants (see \cite{AJG89a} for a
review and `historical notes') Tony Guttmann has given us an invaluable tool which 
over the years has been proved to be by far the best (in terms both of accuracy and 
versatility) method for analysing series. 
In this paper we use long series expansions for self-avoiding polygons and walks
on the honeycomb lattice to test the accuracy of various methods for series analysis. 
For the honeycomb lattice the connective constant, critical exponents and many 
amplitude ratios are known exactly, making it the perfect test-bed for series analysis 
techniques.

The rest of the paper is organised as follows: In section~\ref{sec:th} we give
precise definitions of the models and the properties we investigate and summarise
a number of exact results. Section~\ref{sec:enum} contains a very brief introduction
to the literature describing the algorithms used for the exact enumerations.
In section~\ref{sec:DA} we give a brief introduction to the numerical technique of 
differential approximants and then proceed to analyse the SAP and SAW series 
clearly demonstrating how we can obtain very accurate estimates for the connective
constant and critical exponents. Section~\ref{sec:Ampl} is concerned with
the estimation of amplitudes. Not only do we obtain very accurate estimates 
for the amplitudes, but we also show how an analysis of the asymptotic behaviour
of the series coefficients can be used to gain insight into corrections to scaling.
Finally, in section~\ref{sec:sum} we discuss and summarise our main results.

\section{Definitions and theoretical background \label{sec:th}}   

\begin{figure}
\begin{center}
\includegraphics{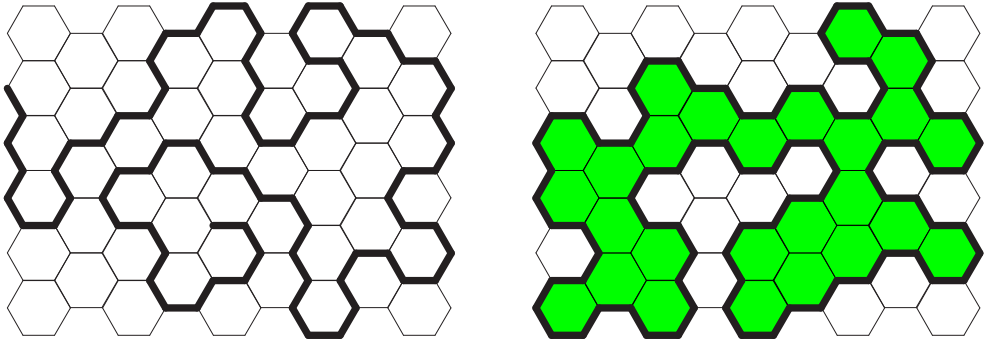}
\end{center}
\caption{\label{fig:example} 
Examples of a self-avoiding walk (left panel) and polygon (right panel) on the
honeycomb lattice.  
}
\end{figure}

An {\em $n$-step self-avoiding walk} $\bm{\omega}$  is 
a sequence of {\em distinct} vertices $\omega_0, \omega_1,\ldots , \omega_n$ 
such that each vertex is a nearest neighbour of it predecessor. SAWs are
considered distinct up to translations of the starting point $\omega_0$.
We shall use the symbol $\bm{\Omega}_n$ to mean the set of all 
SAWs of length $n$. A self-avoiding polygon of length $n$ is an  $n-1$-step SAW
such that $\omega_0$ and $\omega_{n-1}$ are nearest neighbours and a closed loop
can be formed by inserting a single additional step between the two end-points
of the walk. The two models are illustrated in figure~\ref{fig:example}. One is
interested in the number of SAWs and SAPs of length $n$, various metric properties
such as the radius of gyration, and for SAPs one can also ask about the 
area enclosed by the polygon. In this paper we study the following properties:

\begin{itemize}\setlength{\itemsep}{0mm}
\item[(a)] the number of $n$-step self-avoiding walks $c_n$;
\item[(b)] the number of $n$-step self-avoiding polygons  $p_n$;
\item[(c)] the mean-square end-to-end distance of $n$-step SAWs $\ave{R^2_e}_n$;
\item[(d)] the mean-square radius of gyration of $n$-step SAWs $\ave{R^2_g}_n$;
\item[(e)] the mean-square distance of a monomer from the end points of $n$-step 
SAWs $\ave{R^2_m}_n$;
\item[(f)] the mean-square radius of gyration of $n$-step SAPs $\ave{R^2}_n$; and
\item[(g)] the $m^{\rm th}$ moment of the area of  $n$-step SAPs $\ave{a^m}_n$.
\end{itemize}

It is generally believed that the quantities listed above have the asymptotic forms
as $n \to \infty$:

\numparts
\begin{eqnarray}
c_n  &=&  A \mu^n n^{\gamma-1}[1 + o(1)], \label{eq:asympsaw} \\
p_n  &=&  B \mu^n n^{\alpha-3}[1 + o(1)], \label{eq:asympsap}  \\
\ave{R^2_e}_n  &=&  Cn^{2\nu}[1 + o(1)], \label{eq:asympee} \\
\ave{R^2_g}_n  &=&  Dn^{2\nu}[1 + o(1)], \label{eq:asymprg}\\
\ave{R^2_m}_n &=&  En^{2\nu}[1 + o(1)],  \label{eq:asympmd} \\
\ave{R^2}_n   &=&  Fn^{2\nu}[1 + o(1)],   \label{eq:asympsaprg} \\
\ave{a^m}_n   &=&  G^{(m)}n^{2\nu m}[1 + o(1)]. \label{eq:asympmom} 
\end{eqnarray}
\endnumparts 
The critical exponents are believed to be universal in that they only depend
on the dimension of the underlying lattice.  The connective constant
$\mu$ and the critical amplitudes $A$--$G^{(m)}$  vary from lattice to lattice.
In two dimensions the critical exponents $\gamma = 43/32$, 
$\alpha =1/2$ and $\nu = 3/4$ have been predicted exactly, though 
non-rigorously \cite{Nienhuis82a,Nienhuis84a}.
In this work Nienhuis also predicted the exact value of the connective constant
on the honeycomb lattice $\mu=\sqrt{2+\sqrt{2}}$.
When analyzing the series data it is often convenient to use the associated 
generating functions such as
\begin{eqnarray}
\CGf(x)& = &\sum_{n=0} c_n x^n  \sim A(x)(1-\mu x)^{-\gamma}, \label{eq:sawgf} \\
\PGf(x)& = &\sum_{n=0} p_{2n+6 }x^n \sim B(x)(1-\mu^2 x)^{2-\alpha}. \label{eq:sapgf}
\end{eqnarray}
In the polygon generating function we take into account
that SAPs have even length and the smallest one has perimeter 6. 
The SAW (SAP) generating function has a singularity at the critical
point $x=x_c = 1/\mu$ ($x=x_c^2=1/\mu^2$) with critical exponent $-\gamma$ ($2-\alpha$). 

The metric properties for SAWs are defined by,

\begin{eqnarray*}
\ave{R^2_e}_n &= & \frac{1}{c_n} \sum_{\bm{\Omega}_n} (\omega_0 - \omega_n)^2, \\
\ave{R^2_g}_n &= &\frac{1}{c_n} \sum_{\bm{\Omega}_n}\left [ \frac{1}{2(n+1)^2} 
\sum_{i,j=0}^n (\omega_i - \omega_j)^2 \right ], \\
\ave{R^2_m}_n &= &\frac{1}{c_n} \sum_{\bm{\Omega}_n} \left [ \frac{1}{2(n+1)}
\sum_{i=0}^n \left [(\omega_0-\omega_j)^2+(\omega_n-\omega_j)^2 \right ] \right ],
\end{eqnarray*}
\noindent
with a similar definition for the radius of gyration of SAPs.

While the amplitudes are non-universal, there are many universal amplitude
combinations. Any ratio of the metric SAW amplitudes, e.g. $D/C$ and $E/C$, 
is expected to be universal \cite{CS89}. Of particular interest is the linear 
combination \cite{CS89,CPS90} (which we shall call the CSCPS relation)
\begin{equation} \label{eq:CSCPS}
 H \;\equiv\;
   \left( 2 +  \frac{y_t}{y_h} \right)  \frac{D}{C}
   \,-\, 2 \frac{E}{C} \,+\, \frac12,
\end{equation}
where $y_t = 1/\nu$ and $y_h = 1 + \gamma/(2\nu)$. In two dimensions  
Cardy and Saleur \cite{CS89} (as corrected by 
Caracciolo, Pelissetto and Sokal \cite{CPS90}) have predicted, using 
conformal field theory, that $H = 0$. 
Cardy and Guttmann \cite{CG93} proved that $BF=\frac{5}{32\pi^2}\sigma a_0$, where
$\sigma$ is an integer constant such that 
$p_n$ is non-zero when $n$ is divisible by $\sigma$, so $\sigma=2$ for the honeycomb lattice. 
$a_0=3\sqrt{3}/4$ is the area per lattice site on the honeycomb lattice.
Richard, Guttmann and Jensen \cite{RGJ01} conjectured the exact form of
the critical scaling function for self-avoiding polygons and consequently 
showed that the amplitude combinations $G^{(k)}B^{k-1}$ are universal and
predicted their exact values. The exact value for $G^{(1)}=\frac{1}{4\pi}$
had previously been predicted by Cardy \cite{JLC94a}.

The asymptotic form (\ref{eq:asympsaw}) only explicitly gives the leading contribution. 
In general one would expect corrections to scaling so
\begin{equation}
c_n= A\mu^n n^{\gamma-1}\left [1 + \frac{a_1}{n}+\frac{a_2}{n^2}+\ldots
+ \frac{b_0}{n^{\Delta_1}}+\frac{b_1}{n^{\Delta_1+1}}+\frac{b_2}{n^{\Delta_1+2}}+\ldots
\right]
\end{equation}
In addition to ``analytic'' corrections to scaling of the form $a_k/n^k$,
where $k$ is an integer, there are ``non-analytic'' corrections to scaling of the 
form $b_k/n^{\Delta_1+k}$, where the correction-to-scaling exponent $\Delta_1$ 
isn't an integer. In fact one would expect a whole sequence of
correction-to-scaling exponents $\Delta_1 < \Delta_2 \ldots$, which
are both universal and also independent of the observable, that is,
the same for $c_n$, $p_n$, and so on. Furthermore, there should
also be corrections with exponents such as $n\Delta_i+m\Delta_j$, etc.,
with $n$ and $m$ positive integers.  
At least two different theoretical predictions have been made
for the exact value of the leading non-analytic correction-to-scaling exponent:
$\Delta_1 = 3/2$ based on Coulomb-gas arguments \cite{Nienhuis82a,Nienhuis84a} 
and $\Delta_1 = 11/16$ based on conformal-invariance methods \cite{Saleur87a}. 
In a recent paper \cite{CGJPRS} the amplitudes and the correction-to-scaling exponents 
for SAWs on the square and triangular lattices were studied in great detail.
The analysis provided firm numerical evidence that $\Delta_1=3/2$ as predicted by Nienhuis.

\section{Enumerations \label{sec:enum}}

The algorithm we used to enumerate SAPs on the honeycomb lattice is based on
the finite-lattice method devised by Enting \cite{IGE80e} in his pioneering work,
which contains a detailed description of the original approach for enumerating
SAPs on the square lattice. A major enhancement, resulting in an exponentially more 
efficient algorithm, is described in some detail in \cite{JG99} while details of the changes 
required to enumerate area-moments and the radius of gyration can be found in \cite{IJ00a}.
A very efficient parallel implementation is described in \cite{IJ03a}. The
generalisation to enumerations of SAWs is straight forward as shown in \cite{IJ04a}.
An implementation of the basic SAP enumeration algorithm on the honeycomb lattice 
can be found in \cite{EG89a}. Most of the enhancements we made to the square
lattice case can also be readily implemented on the honeycomb lattice. The only
slightly tricky part is the calculation of metric properties (though the changes are very 
similar to those required for the triangular lattice \cite{IJ04d}). 

Using the a parallel version of our honeycomb lattice algorithms we have counted the number of 
self-avoiding walks and polygons to length 105 and 158, respectively. For self-avoiding walks 
to length 96 we also calculate series for the metric properties of mean-square end-to-end 
distance, mean-square radius of gyration and the mean-square distance of a 
monomer from the end points. In fact the algorithm calculates the metric generating
functions with coefficients $c_n\ave{R^2_e}_n$, $n^2 c_n\ave{R^2_g}_n$, and 
$nc_n\ave{R^2_m}_n$, respectively, the advantage being that these quantities are integer valued.
For self-avoiding polygons to length 140 we calculate
series for the mean-square radius of gyration and the first 10 moments of the 
area. Again we actually calculate the series with integer coefficients
$8n^2p_n\ave{R^2}_n$ and $p_n\ave{a^k}_n$.

\section{Differential approximants \label{sec:DA}}

The majority of interesting models in statistical mechanics and combinatorics 
have generating functions with regular singular points such as those indicated in
(\ref{eq:sawgf}) and  (\ref{eq:sapgf}). The fundamental problem of series
analysis is: Given a {\em finite} number of terms in the series expansion
for a function $F(x)$ what can one say about the singular behaviour which 
after all is a property of the {\em infinite} series. Without a doubt the
best series analysis technique when it comes to locating singularities and estimating
the associated critical exponents is differential approximants (see \cite{AJG89a}
for a comprehensive review of differential approximants and other techniques for series 
analysis). The basic idea is to approximate the function $F(x)$ by solutions
to differential equations with polynomial coefficients. The singular behaviour
of such ODEs is much studied (see \cite{Ince}) and the singular points and
exponents are easily calculated.

A $K^{th}$-order differential approximant (DA) to a function $F(x)$  is formed by matching 
the coefficients in the polynomials $Q_i(x)$ and $P(x)$ of degree $N_i$ and $L$, respective,
so that (one) of the formal solutions to the inhomogeneous differential equation
$$
\sum_{i=0}^K Q_{i}(x)(x\frac{{\rm d}}{{\rm d}x})^i \tilde{F}(x) = P(x)
$$
agrees with the first $M=L+\sum_i (N_i+1)$ series coefficients of $F$.
Singularities of $F(x)$ are approximated by the zeros $x_i$ of $Q_K(x)$ and the 
associated critical exponent $\lambda_i$ is estimated from the indicial equation. 
If there is only a single root at $x_i$  this is just
$$
\lambda_i=K-1-\frac{Q_{K-1}(x_i)}{x_iQ_K ' (x_i)}.
$$
The physical critical point is the first singularity on the positive real axis.

In order to locate the singularities of the series in a systematic fashion we used the
following procedure: We calculate all $[L;N_0,N_1,N_2]$ and $[L;N_0,N_1,N_2,N_3]$ second- and 
third-order inhomogeneous differential approximants with $|N_i - N_j| \leq 2$, that is 
the degrees of the polynomials $Q_i$ differ by at most 2. In addition we demand that the
total number of terms used by the DA is at least $N_{\rm max}-10$, where  $N_{\rm max}$
is the total number of terms available in the series. Each approximant yields 
$N_K$ possible singularities and associated exponents from the $N_K$ zeroes of $Q_K(x)$ 
(many of these are of course not actual singularities of the series but merely spurious zeros.) 
Next these zeroes are sorted into equivalence classes by the criterion that they lie at most a 
distance $2^k$ apart. An equivalence class is accepted as a singularity if it appears in more than 
75\% of the total number of approximants, and an estimate for the singularity and exponent is 
obtained by averaging over the included approximants (the spread among the approximants is also 
calculated). The calculation was then repeated for $k-1$, $k-2$, $\ldots$ until a minimal value 
of 8 was reached. To avoid outputting well-converged singularities at every level, once an 
equivalence class has been accepted, the data used in the estimate is discarded, 
and the subsequent analysis is carried out on the remaining data only. One advantage of this 
method is that spurious outliers, some of which will almost always be present when so many 
approximants are generated, are discarded systematically and automatically.

\subsection{The polygon series  \label{sec:DAsap}}

First we apply our differential approximant analysis to the self-avoiding polygon
generating function. In table~\ref{tab:sapDA} we have listed the estimates for the critical 
point $x_c^2$ and exponent $2-\alpha$ obtained from second- and third-order DAs. 
We note that all the estimates are in perfect agreement (surely a best case scenario)
in that within `error-bars' they take the same value. From this we arrive at the  
estimate $x_c^2=0.2928932186(5)$ and $2-\alpha=1.5000004(10)$, where the error-bars 
reflect the spread among the estimates and the individual error-bars (note that DA estimates 
{\em are not} statistically independent so the final error-bars exceed the individual ones). 
The final estimates are in perfect agreement with the conjectured exact
values $x_c^2=1/\mu^2=1/(2+\sqrt{2})= 0.292893218813\ldots$ and $2-\alpha=3/2$.

\begin{table}
\caption{\label{tab:sapDA} 
Critical point and exponent estimates for self-avoiding polygons.}
\begin{indented}
\item[]\begin{tabular}{lllll} \br
 $L$   &  \multicolumn{2}{c}{Second order DA} & 
       \multicolumn{2}{c}{Third order DA} \\ 
\mr
    &  \multicolumn{1}{c}{$x_c^2$} & \multicolumn{1}{c}{$2-\alpha$} & 
      \multicolumn{1}{c}{$x_c^2$} & \multicolumn{1}{c}{$2-\alpha$} \\ 
\mr
 0 & 0.29289321854(19)& 1.50000065(41) &  0.29289321865(12)& 1.50000040(28) \\
 5 & 0.29289321875(21)& 1.50000010(59) &  0.29289321852(48)& 1.50000041(99) \\
10 & 0.29289321855(23)& 1.50000060(48) &  0.29289321878(32)& 1.49999999(97) \\
15 & 0.29289321859(19)& 1.50000054(43) &  0.29289321861(37)& 1.50000035(67) \\
20 & 0.29289321866(15)& 1.50000038(33) &  0.29289321860(21)& 1.50000049(43) \\
\br
\end{tabular}
\end{indented}
\end{table}

\begin{figure}
\begin{center}
\includegraphics[scale=0.55]{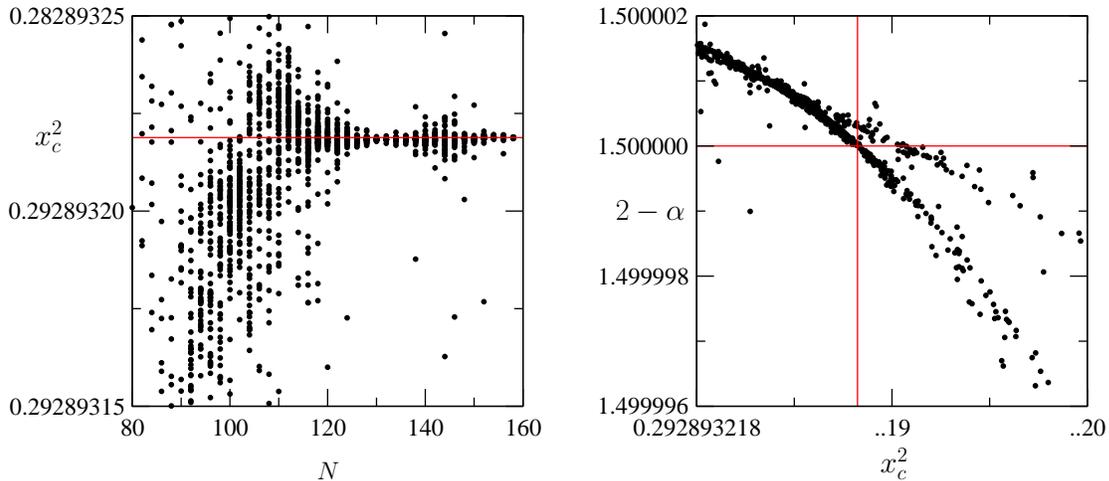}
\end{center}
\caption{\label{fig:sapDA}
Plot of estimates from third order differential approximants 
for $x_c^2$ vs. the highest order term used and $2-\alpha$ vs. $x_c^2$. The straight lines
are the exact predictions.
}
\end{figure}

Before proceeding we will consider possible sources of systematic errors. First and
foremost the possibility that the estimates might display a systematic drift as the number
of terms used is increased and secondly the possibility of numerical errors. The latter
possibility is quickly dismissed. The calculations were performed using 128-bit real numbers.
The estimates from a few approximants were compared to values obtained using MAPLE with  
100 digits accuracy and this clearly showed that the program was numerically stable and rounding 
errors were negligible. In order to address the possibility of systematic drift and lack of 
convergence to the true critical values we refer to figure~\ref{fig:sapDA} (this is probably
not really necessary in this case but we include the analysis here in order to present 
the general method). In the left panel of figure~\ref{fig:sapDA} we have plotted the 
estimates  from third-order DAs for $x_c^2$ vs. the highest order term $N$ used by the DA.
Each dot in the figure is an estimate obtained from a specific approximant. As can be
seen the estimates clearly settle down to the conjectured exact value (solid line) as
$N$ is increased and there is little to no evidence of any systematic drift at large $N$.
One curious aspect though is the widening of the spread in the estimates around $N=140$.
We have no explanation for this behaviour but it could quite possibly be caused by just a few 
`spurious' approximants.
In the right panel we show the variation in the exponent estimates with the critical point
estimates. We notice that the `curve' traced out by the estimates pass through the intersection
of the lines given by the exact values. We have not been able to determine the reason for the
apparent branching into two parts. However, we note that the lower `branch' contain many
more approximants than the upper one.

\begin{figure}
\begin{center}
\includegraphics[scale=0.55]{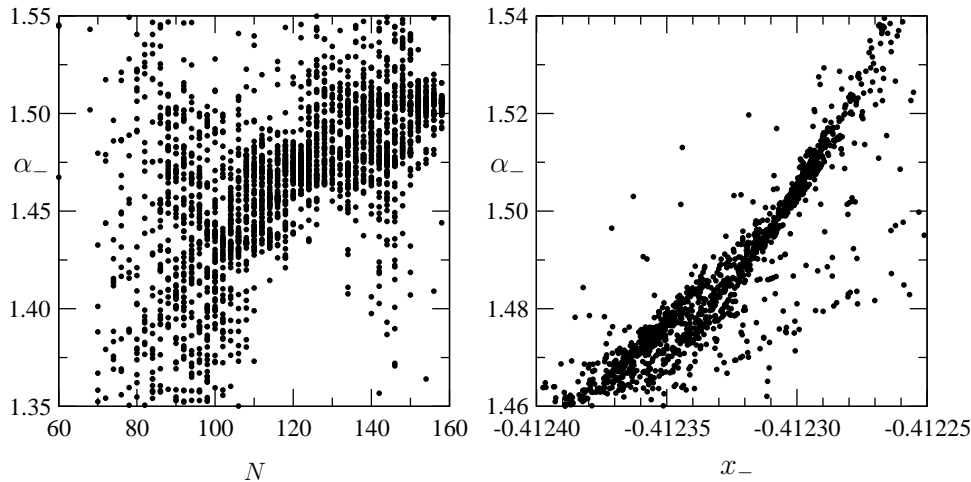}
\end{center}
\caption{\label{fig:sapneg}
Plot of estimates from third order differential approximants for the location $x_-$
of the non-physical singularity and the associated exponent $\alpha_-$. The left
shows $\alpha_-$ vs. the highest order term used and right panel $\alpha_-$ vs. $x_-$.}
\end{figure}

The differential approximant analysis can also be used to find possible non-physical
singularities of the generating function. Averaging over the estimates from the DAs shows
that there is an additional non-physical singularity on the negative $x$-axis at 
$x=x_-=-1/\mu_-^2 =-0.41230(2)$, where the associated critical exponent $\alpha_-$ has a value 
consistent with the exact value $\alpha_-=3/2$. In the left panel of figure~\ref{fig:sapneg} 
we have plotted  $\alpha_-$ vs. the highest order term used by the DAs and we clearly
see the convergence to a value consistent with $\alpha_-=3/2$. If we take this value
as being exact we can get a refined estimate of $x_-$ from the plot in the right panel
of  figure~\ref{fig:sapneg}, where we notice that the estimates for $\alpha_-$ cross
the value $3/2$ for $x_-=-0.412305(5)$ which we take as our final estimate. From
this we then get $\mu_- = 1.557366(10)$.

\subsection{The walk series  \label{sec:DAsaw}}

Next we apply the differential approximant analysis to the self-avoiding walk
generating function. In table~\ref{tab:sawDA} we have listed the estimates for the critical 
point $x_c$ and exponent $\gamma$ obtained from second- and third-order DAs. 
Firstly, we note that estimates are about an order of magnitude less accurate than in 
the polygon case. Secondly, there are now small but nevertheless seemingly systematic 
differences between the second- and third order DAs (in particular the second-order
homogeneous ($L=0$) approximants are much less accurate than the other cases).
On general theoretical grounds one would expect higher-order inhomogeneous approximants
to be better in that they can accommodate more complicated functional behaviour.
So based mainly on the third-order DAs we finally estimate that $0.541196102(4)$ and
$\gamma=1.343758(8)$. This is consistent with the exact values 
$x_c=1/\mu=1/\sqrt{2+\sqrt{2}}=0.541196100146\ldots$ and $\gamma=43/32=1.34375$, though
the central estimates for $\gamma$ are systematically a bit too high (and the second-order
DAs are worse).

\begin{table}
\caption{\label{tab:sawDA} 
Critical point and exponent estimates for self-avoiding walks.}
\begin{indented}
\item[]\begin{tabular}{lllll} \br
 $L$   &  \multicolumn{2}{c}{Second order DA} & 
       \multicolumn{2}{c}{Third order DA} \\ 
\mr
    &  \multicolumn{1}{c}{$x_c$} & \multicolumn{1}{c}{$\gamma$} & 
      \multicolumn{1}{c}{$x_2$} & \multicolumn{1}{c}{$\gamma$} \\ 
\mr
 0 & 0.541196097(19)  & 1.34360(36)   & 0.5411961075(19)  & 1.3437685(54) \\
 5 & 0.5411961066(10) & 1.343770(18)  & 0.5411961025(10)  & 1.3437583(19) \\
10 & 0.5411961065(12) & 1.3437669(53) & 0.54119610266(91) & 1.3437584(20) \\
15 & 0.5411961069(16) & 1.343776(68)  & 0.5411961011(17)  & 1.3437551(38) \\
20 & 0.5411961059(21) & 1.3437646(29) & 0.5411961022(26)  & 1.3437580(59) \\
\br
\end{tabular}
\end{indented}
\end{table}

In addition there is a singularity on the negative $x$-axis at $x=-x_c$ with
a critical exponent consistent with the value $1/2$, and a pair of singularities
at $x=\pm  0.64215(15){\rm i} $ with an exponent which is likely to equal $1/2$
(note that  the value $0.64215(15)$ is consistent with $1/\mu_-$). These results
help to at least partly explain why the walk series is more difficult to analyse than
the polygon series. The walk series has more non-physical singularities and one of
these (at $x=-x_c$) is closer to the origin than the  non-physical singularity
of the polygon series. Furthermore, as argued and confirmed
numerically in the next section, the walk series has non-analytical corrections to
scaling whereas the polygon series has only analytical corrections. All of these 
effects conspire to make the walk series much harder to analyse and it is indeed
a great testament to the method of differential approximants that the analysis given
above yields such accurate estimates despite all these complicating factors.

\begin{figure}
\begin{center}
\includegraphics[scale=0.55]{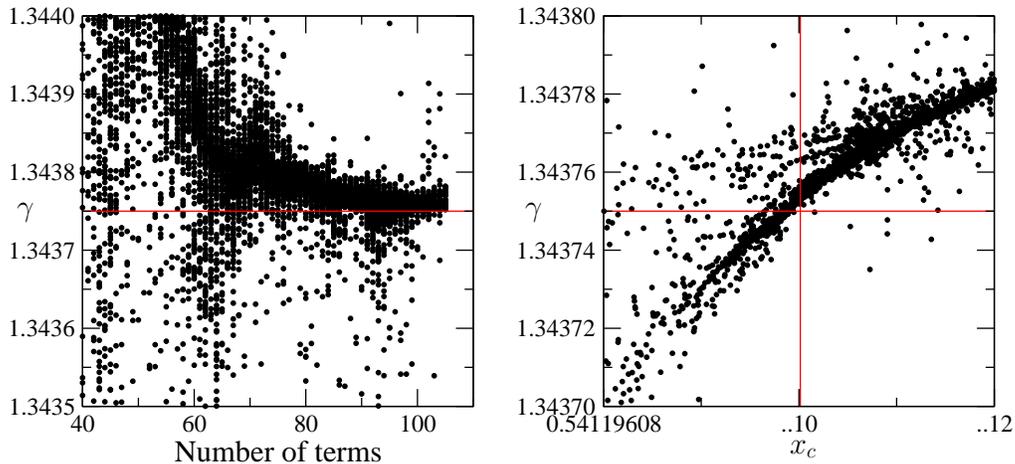}
\end{center}
\caption{\label{fig:sawDA}
Plot of estimates from third order differential approximants for $\gamma$ vs. the number 
of terms used by the DA and $\gamma$ vs. $x_c$. The straight lines are the exact predictions.
}
\end{figure}

In figure~\ref{fig:sawDA} we have plotted estimates for $\gamma$, 
obtained from third-order DAs, against the number of terms used by the DA (left panel)
and against estimates for the critical point $x_c$ (right panel). The estimates for 
$\gamma$ display some rather curious and unexpected variations with the number of terms.
Early on (around 80 terms) the estimates seems to settle at a value above the
exact result. A little later the estimates start trending downwards so that around 95 terms 
they are in excellent agreement with the exact value. However, the estimates then unexpectedly 
start trending upwards again so that with more than a 100 terms the agreement with the
exact result is only marginal. We also notice that in the right panel estimates
for $\gamma$ vs. $x_c$ happen to just miss the intersection between the lines marking the
exact predictions. This behaviour is curious and we have no ready explanation for
it other than once again drawing attention to the quite complicated functional form of the
generating function. The discrepancies between the estimates and exact values is marginal 
and certainly not significant enough to raise questions about the correctness of the 
theoretical predictions.

\subsection{Metric properties \label{sec:DAmetric}}

Finally, we briefly turn our attention to the series for the metric properties of
SAPs and SAWs. We actually study the metric generating functions with integer coefficients
$8n^2p_n\ave{R^2}_n$, $c_n \ave{R^2_e}_n$, $n^2 c_n\ave{R^2_g}_n$, and $nc_n\ave{R^2_m}_n$, 
which have critical exponents $-(\alpha-2\nu)=-2$, $-(\gamma+2\nu)=-91/32=-2.84375$,
$-(\gamma+2\nu+2)=-155/32=-4.84375$, and $-(\gamma+\nu+1)=-123/32=-3.84375$, respectively 
(and as usual the polygon series use only the even terms). In table~\ref{tab:metric} we list 
the estimates obtained for the critical point and exponents using averages over third-order DAs. 
The exponent estimates from the SAP series are consistent with the expected value
confirming $\nu=3/4$ as are the estimates from the SAW series.  The only possible
exception is the end-to-end distance series where the estimates for both
$x_c$ and the exponent are systematically a little to high. However, the discrepancy
is not very large and probably not significant.

\begin{table}
\caption{\label{tab:metric} 
Critical point and exponent estimates for metric properties of SAPs and SAWs.}
\begin{indented}
\item[]\begin{tabular}{lllll} \br
 $L$   &  \multicolumn{2}{c}{SAP radius of gyration} & 
       \multicolumn{2}{c}{SAW end-to-end distance} \\ 
\mr
    &  \multicolumn{1}{c}{$x_c^2$} & \multicolumn{1}{c}{$\alpha+2\nu$} & 
      \multicolumn{1}{c}{$x_c$} & \multicolumn{1}{c}{$\gamma+2\nu$} \\ 
\mr
 0 & 0.292893246(10)  & 2.000176(35) &  0.5411961141(14) & 2.8438094(28) \\
 5 & 0.2928932440(70) & 2.000169(24) &  0.5411961136(31) & 2.8438080(64)\\
10 & 0.292893245(24)  & 2.000166(92) &  0.5411961124(37) & 2.8438054(82)\\
15 & 0.292893235(61)  & 2.00008(27)  &  0.5411961133(33) & 2.8438072(66)\\
20 & 0.292893262(42)  & 2.00019(11)  &  0.5411961113(25) & 2.8438031(59) \\
\mr    &  \multicolumn{2}{c}{SAW radius of gyration} & 
       \multicolumn{2}{c}{SAW distance from end-point} \\ 
\mr
    &  \multicolumn{1}{c}{$x_c$} & \multicolumn{1}{c}{$\gamma+2\nu+2$} & 
      \multicolumn{1}{c}{$x_c$} & \multicolumn{1}{c}{$\gamma+\nu+1$} \\ 
\mr
 0 &  0.541196111(22)  & 4.843788(47) & 0.5411961013(28) & 3.8437852(95) \\
 5 &  0.541196115(12)  & 4.843806(19) & 0.5411961014(21) & 3.8437843(90)\\
10 &  0.5411961041(91) & 4.843789(21) & 0.5411961033(52) & 3.843789(11)\\
15 &  0.5411961021(77) & 4.843784(21) & 0.5411961064(75) & 3.843794(22)\\
20 &  0.5411961040(49) & 4.843794(10) & 0.5411961049(40) & 3.8437954(75)\\
\br
\end{tabular}
\end{indented}
\end{table}

\section{Amplitude estimates \label{sec:Ampl}}

Now that the exact values of $\mu$ and the exponents have been confirmed 
we turn our attention to the ``fine structure'' of the asymptotic form of the
coefficients. In particular we are interested in obtaining accurate
estimates for the leading critical amplitudes such as $A$ and $B$. Our method
of analysis consists in fitting the coefficients to an assumed asymptotic form.
Generally we must include a number of terms in order to account
for the behaviour of the generating function at the physical singularity,
the non-physical singularities as well as sub-dominant corrections to
the leading order behaviour. As we hope to demonstrate, this method of
analysis can not only yield accurate amplitude estimates, but it is often possible
to clearly demonstrate which corrections to scaling are present.

Before proceeding with the analysis we briefly consider the kind of terms which occur
in the generating functions, and how they influence the asymptotic behaviour of the series 
coefficients. At the most basic level a function $G(x)$ with a power-law singularity

\begin{equation}\label{eg:Gbasic}
G(x) =\sum_n g_n x^n \sim A(x)(1-\mu x)^{-\xi},
\end{equation}
where $A(x)$ is  an analytic function at $x=x_c$, gives rise to the asymptotic form 
of the coefficients

\begin{equation}\label{eq:CoAbasic}
g_n \sim \mu^n n^{\xi-1} \left [ \tilde{A}+ \sum_{i\geq 1} a_i/n^i \right ],
\end{equation}
that is we get a dominant exponential growth given by $\mu$, modified by a sub-dominant 
term given by the critical exponent followed by analytic corrections. The amplitude $\tilde{A}$
is related to the function $A(x)$ in (\ref{eg:Gbasic}) via the relation 
$\tilde{A}=A(1/\mu)/\Gamma(\xi)$. If $G(x)$ has a non-analytic correction to scaling such as
\begin{equation}\label{eq:Gcorr}
G(x) =\sum_n g_n x^n \sim (1-\mu x)^{-\xi}\left [A(x)+B(x)(1-\mu x)^{\Delta}\right ],
\end{equation}
we get the more complicated form

\begin{equation}\label{eq:CoAcorr}
g_n \sim \mu^n n^{\xi-1} \left [ \tilde{A}+ \sum_{i\geq 1} a_i/n^i+ \sum_{i\geq 0} b_i/n^{\Delta+i} \right ].
\end{equation}
A singularity on the negative $x$-axis $\propto (1+\mu_- x)^{-\eta}$ leads to additional corrections
of the form

\begin{equation}\label{eq:CoAneg}
\sim (-1)^n \mu_-^n n^{\eta-1} \sum_{i\geq 0} c_i/n^i.
\end{equation}
Singularities in the complex plane are more complicated. However, a pair of singularities 
in the complex axis at $\pm {\rm i}/\tau$, that is a term of the form $D(x)(1+\tau^2 x^2)^{-\eta}$, 
generally results in coefficients that  change sign according to a $++--$ pattern.
This can be accommodated by terms of the form

\begin{equation}\label{eq:CoAcomp}
\sim (-1)^{\lfloor n/2 \rfloor}\tau^n n^{\eta-1} \sum_{i\geq 0} d_i/n^i.
\end{equation}

All of these possible contribution must then be put together in an assumed asymptotic
expansion for the coefficients $g_n$ and we obtain estimates for the unknown amplitudes
by directly fitting $g_n$ to the assumed form. That is we take a sub-sequence of terms
$\{g_n,g_{n-1},\ldots,g_{n-k}\}$, plug into the assumed form and solve the 
$k+1$ linear equations to obtain estimates for the first few amplitudes.
As we shall demonstrate below this allows us to probe the asymptotic form. 

\subsection{Estimating the polygon amplitude $B$ \label{sec:Bampl}}

\begin{figure}
\begin{center}
\includegraphics[scale=0.72]{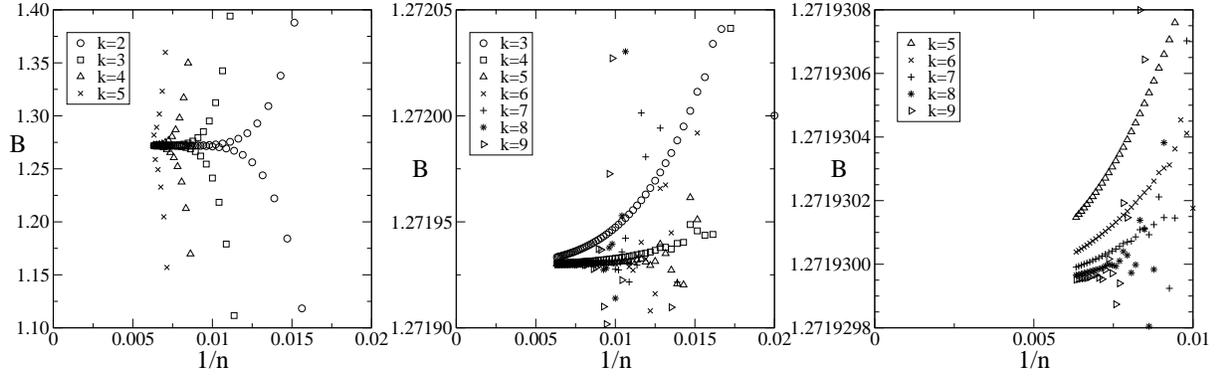}
\end{center}
\caption{\label{fig:sapampl}
Plots of fits for the self-avoiding polygon amplitude $B$ using in the left panel the asymptotic 
form (\protect{\ref{eq:Bnoneg}}) which ignores the singularity at $x=x_-$, and in the middle panel
the asymptotic form (\protect{\ref{eq:Bneg}}) which includes the singularity at $x=x_-$.
The right panel gives a closer look at the data form the middle panel.
}\end{figure}

The asymptotic form of the coefficients $p_n$ of the generating function of square and
triangular lattice SAPs has been studied in detail previously \cite{CG96,JG99,IJ03a,IJ04d}. 
There is now clear numerical evidence that the leading correction-to-scaling exponent for SAPs 
and SAWs is $\Delta_1=3/2$, as predicted by Nienhuis \cite{Nienhuis82a,Nienhuis84a}.
As argued in \cite{CG96} this leading correction term combined with the $2-\alpha=3/2$ 
term of the SAP generating function produces an {\em analytic} background term as
can be seen from eq.~(\ref{eq:Gcorr}). Indeed in the previous analysis of SAPs there was 
no sign of non-analytic corrections-to-scaling to the generating function (a strong indirect 
argument that the leading correction-to-scaling exponent must be half-integer valued). At first 
we ignore the singularity at $x_-$ (since $|x_-| > x_c^2$ it is exponentially suppressed) and
obtain estimates for $B$ by fitting $p_n$ to the form 
\begin{equation}\label{eq:Bnoneg}
p_n = \mu^n n^{-5/2} \left [ B+ \sum_{i=1}^k a_i/n^i \right ].
\end{equation}
That is we take a sub-sequence of terms
$\{p_n,p_{n-2},\ldots,p_{n-2k}\}$ ($n$ even), plug into the formula above and solve the 
$k+1$ linear equations to obtain estimates for the amplitudes.
It is then advantageous to plot estimates for the leading amplitude $B$ against $1/n$ for several
values of $k$. The results are plotted in the left panel of figure~\ref{fig:sapampl}.
Obviously the amplitude estimates are not well behaved and display clear parity effects.
So clearly we can't just ignore the singularity at $x_-$ (which gives rise to such effects)
and we thus try fitting to the more general form

\begin{equation}\label{eq:Bneg}
p_n = \mu^n n^{-5/2} \left [ B+ \sum_{i= 1}^k a_i/n^i \right ]+
(-1)^{n/2}\mu_-^n n^{-5/2} \sum_{i=0}^k b_i/n^i.
\end{equation}
The results from these fits are shown in the middle panel of  figure~\ref{fig:sapampl}.
Now we clearly have very well behaved estimates (note the significant change of scale 
along the $y$-axis from the left to the middle panel). In the right panel we take a more
detailed look at the data and from the plot we estimate that $B=1.2719299(1)$.
We  notice that as more and more correction terms are added ($k$ is increased)
the plots of the amplitude estimates exhibits less curvature and the slope become less steep. 
This is very strong evidence that (\ref{eq:Bneg})  indeed is the correct 
asymptotic form of $p_n$.

\subsection{Estimating the walk amplitude $A$  \label{sec:Aampl}}

\begin{figure}[t]
\begin{center}
\includegraphics[scale=0.8]{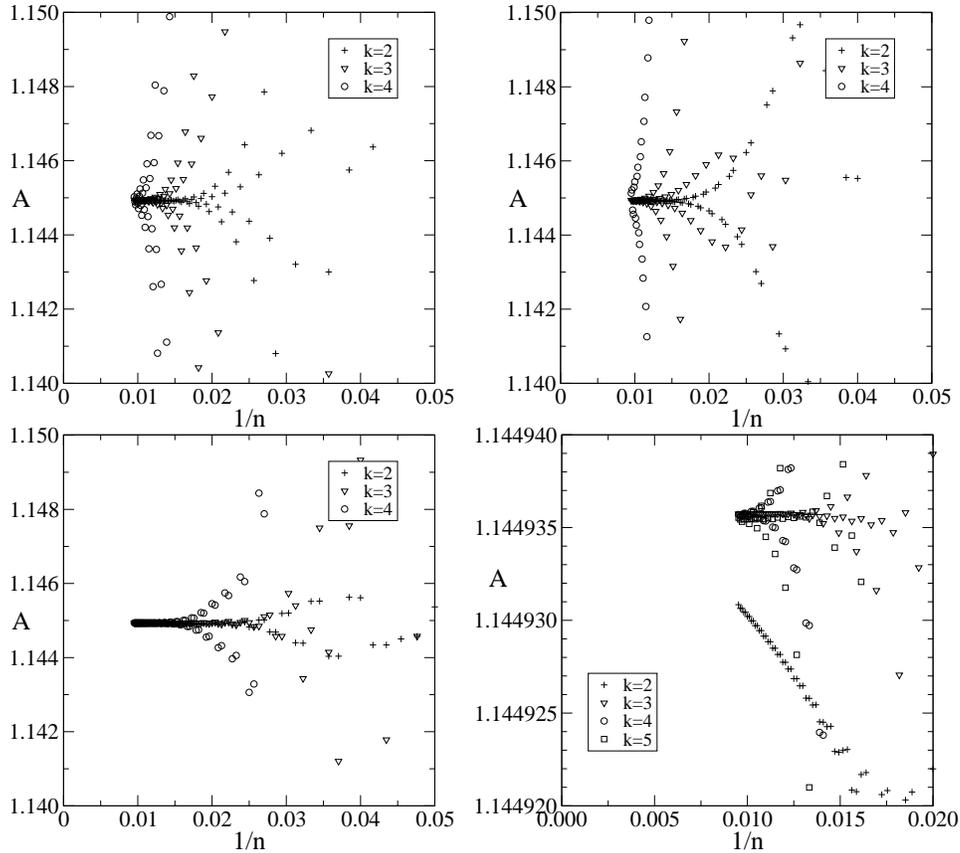}
\end{center}
\caption{\label{fig:sawampl}
Plots of fits for the self-avoiding walk amplitude $A$ using different asymptotic forms.
In the top left panel we show plots using the form (\protect{\ref{eq:Aneg}}) which ignores
the complex singularity. In the top right panel we include the complex singularity via the 
form (\protect{\ref{eq:Acomp}}) while the bottom left panel includes the complex singularity 
via the alternative form (\protect{\ref{eq:Afinal}}). The bottom right panel is a more detailed
look at this latter case.
}
\end{figure}

From the differential approximant analysis we found that the walk generating function has 
non-physical singularities at $x=-x_c$ and $x=\pm \rmi/\mu_-$. In addition we expect from 
Nienhuis's results (confirmed by extensive numerical work \cite{CGJPRS}) a non-analytic
correction-to-scaling term with exponent $\Delta=3/2$, and since $\gamma=43/32$ this
correction term does not vanish in the walk case. Ignoring for the moment the pair of
complex singularities  we first try  with to fit to the asymptotic form 

\begin{equation}\label{eq:Aneg}
c_n = \mu^n n^{11/32} \left [ A + \sum_{i=2}^k a_i/n^{i/2} \right ]+
(-1)^{n}\mu^n n^{-3/2} \sum_{i=0}^k b_i/n^i,
\end{equation}
where the first sum starts at $i=2$ because the leading correction is analytic.
The resulting estimates for the leading amplitude $A$ are plotted in the top left panel
of figure~\ref{fig:sawampl}. Clearly these amplitude estimates are not well behaved 
so we better not ignore the complex pair of singularities!
Therefore we try again with the asymptotic form 

\begin{equation}\label{eq:Acomp}
c_n = \mu^n n^{11/32} \left [ A + \sum_{i= 2}^k a_i/n^{i/2} \right ]+
(-1)^{n}\mu^n n^{-3/2} \sum_{i=0}^k b_i/n^i+
(-1)^{\lfloor n/2 \rfloor}\mu_-^n n^{-3/2} \sum_{i=0}^k c_i/n^i.
\end{equation}
The estimates for the leading amplitude $A$ are plotted in the top right panel
in figure~\ref{fig:sawampl}. These amplitude estimates are not well behaved either
and something is not quite right. Next we try to change the way we included the complex
pair of singularities. In (\ref{eq:Acomp}) we have assumed that all terms arising
from the complex singularity have exactly the same sign-pattern. However, if we assume 
that the analytic correction terms  arise from a functional form such as 
$D(x)(1+\mu_-^2 x^2)$, where $D(x)$ is a analytic, then the analytic correction terms
would actually have a {\em shifted} sign-pattern.
We therefore try fitting to the slightly modified asymptotic form 

\begin{equation}\label{eq:Afinal}
c_n = \mu^n n^{11/32} \left [ A + \sum_{i= 2}^k a_i/n^{i/2} \right ]+
(-1)^{n}\mu^n n^{-3/2} \sum_{i=0}^k b_i/n^i+
\mu_-^n n^{-3/2} \sum_{i=0}^k (-1)^{\lfloor (n+i)/2\rfloor }c_i/n^i,
\end{equation}
where it should be noted that all we have done is change the way we include
the terms from the complex singularities so as to shift the sign-pattern by
a unit as $i$ is increased. The new estimates for the leading amplitude are
plotted in the bottom left panel of figure~\ref{fig:sawampl} and quite
clearly the convergence is now very much improved. In the bottom right panel
of figure~\ref{fig:sawampl} we show a much more detailed look at the data and
from this plot we can estimate that $ A=1.1449355(5)$.

\subsection{The correction-to-scaling exponent \label{sec:corr}}

\begin{figure}
\begin{center}
\includegraphics[scale=0.7]{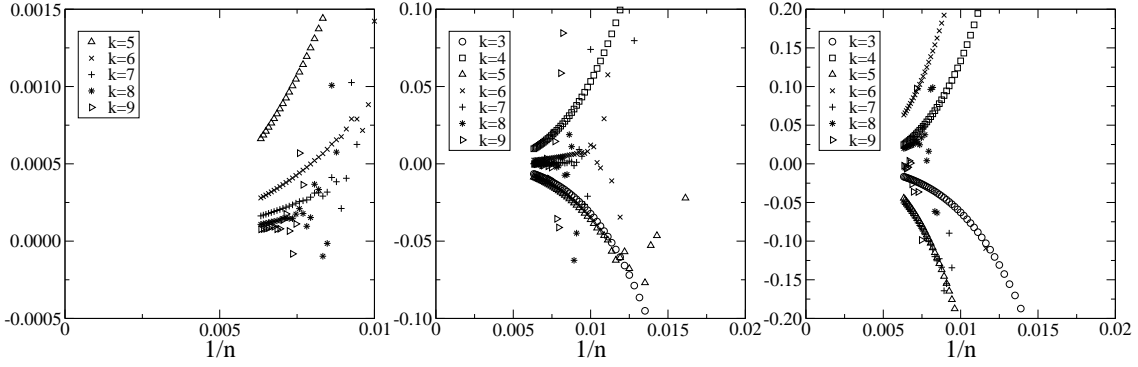}
\end{center}
\caption{\label{fig:sapcorr}
Plots of estimates for the amplitude of the term $1/n^{\Delta_1}$. The left panel
shows results from fits to the form (\protect{\ref{eq:Bcoorfit}}) where only the leading
order term $1/n^{\Delta_1}$ is included (as well as analytical corrections). In the middle
panel additional terms 
of the form $1/n^{\Delta_1+i}$ are included and in the right panel terms like $1/n^{m\Delta_1+i}$
are included.
}
\end{figure}

In this section we shall briefly show how the method of direct fitting 
can be used to differentiate between various possible values for the
leading correction-to-scaling exponent $\Delta_1$ (recall the two
theoretical prediction $\Delta_1=3/2$ by Nienhuis and $\Delta_1=11/16$ by
Saleur). As already stated there is now firm evidence from previous work
that the Nienhuis result is correct. Here we shall present further
evidence. Different values for $\Delta_1$ leads to different assumed
asymptotic forms for the coefficients. For the SAP series we argued that 
a value $\Delta_1=3/2$ (or indeed any half-integer value) would result only 
in {\em analytic} corrections to the generating function and thus that $p_n$ 
asymptotically would be given by (\ref{eq:Bneg}). If on the other
hand we have a generic value for $\Delta_1$ we would get
\begin{equation}\label{eq:Bcoor}
p_n = \mu^n n^{-5/2} \left [ B+ \sum_{i= 1}^k a_i/n^i+ \sum_{i= 0}^k b_i/n^{\Delta_1+i} \right ]+
(-1)^{n/2}\mu_-^n n^{-5/2} \sum_{i=0}^k c_i/n^i.
\end{equation}
Fitting to this form we can then estimate the amplitude $b_0$ of the term $1/n^{\Delta_1}$.
We would expect that if we used a manifestly incorrect value for $\Delta_1$ then
$b_0$ should vanish asymptotically thus demonstrating that this term is really absent from
(\ref{eq:Bcoor}). So we tried fitting to this form using the value $\Delta_1=11/16$.
More precisely we fit to the generic form
\begin{equation}\label{eq:Bcoorfit}
p_n = \mu^n n^{-5/2}\sum_{i=0}^k a_i/n^{\alpha_i}+ (-1)^{n/2}\mu_-^n n^{-5/2} \sum_{i=0}^k b_i/n^i.
\end{equation}
In the first instance we include only the leading term arising from $\Delta_1$,
that is we use the sequence of exponents  $\alpha_i=\{0, 11/16, 1, 2, 3, \ldots \}$.
We also fit to a form in which the additional analytical corrections arising from   $\Delta_1$
are included leading to the sequence of exponents
$\alpha_i=\{ 0, 11/16, 1, 27/16, 2, 33/16, 3, 49/16, \ldots\}$. 
As stated in Section~\ref{sec:th} more generally one would also expect terms of the form 
$1/n^{m\Delta_1+i}$ with $m$ a non-negative integer. This leads to fits to the form given above
but with $\alpha_i = \{0,0.6875, 1, 1.375, 1.6875, 2, 2.0625,  2.375, 2.6875, 2.75, 3 \ldots\}$.
The estimates of the amplitude of the term $1/n^{\Delta_1}$ as obtained from fits to
these forms are shown in figure~\ref{fig:sapcorr}. As can be seen from the left panel,
where we fit to the first case scenario, the amplitude clearly seems to converge to 0,
which would indicate the absence of this term in the asymptotic expansion for $p_n$.
In the middle and right panels we show the results from fits to the more general forms. Again the
estimates are consistent with the amplitude being 0. Though in this case the evidence
is not quite as convincing. This is however not really surprising given that the
incorrect value $\Delta_1=11/16$ gives rise to a plethora of absent terms which will
tend to greatly obscure the true asymptotic behaviour.

\subsection{Amplitude ratios $D/C$ and $E/C$}

From fits to the coefficients in the metric series we find 
$AC=1.0141(1)$, $AD=0.14225(5)$ and $AE=0.4458(1)$ and thus the ratios are

$$D/C=0.14027(6) \mbox{\hspace{10mm} and\hspace{10mm}} E/C=0.43960(15)$$
$D/C$ and $E/C$ can also be estimated directly from the relevant quotient sequence,
e.g. $r_n =  \ave{R^2_g}_n/\ave{R^2_e}_n$, using the following method due to 
Owczarek {\em et al.} \cite{OPBG}:  Given a sequence of the form 
$g_n \sim g_{\infty}(1 + b/n + \ldots)$, we
construct a new sequence $\{h_n\}$ defined by $h_n = \prod_{m=1}^n g_m$.
The associated generating function then has the behaviour
$\sum h_n x^n \sim (1 - g_{\infty} x)^{-(1+b)}$, and we can now
estimate $g_{\infty}$ form a differential approximant analysis. 
In this way, we obtained the estimates
$$D/C = 0.1403001(2) \mbox{\hspace{10mm} and\hspace{10mm}} E/C = 0.439635(1)$$
These amplitude estimates leads to a high precision confirmation of the
CSCPS relation  $H=0.000003(13)$.

\subsection{Amplitude combination $BF$}

\begin{figure}
\begin{center}
\includegraphics[scale=0.45]{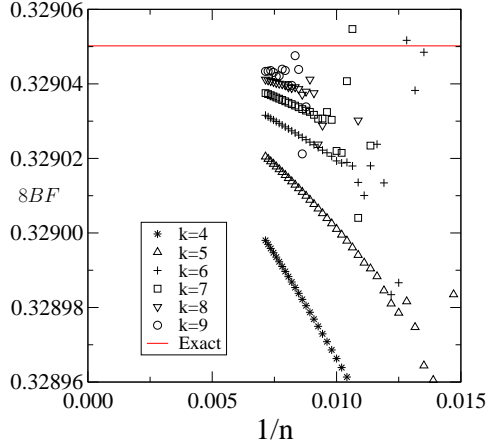}
\end{center}
\caption{\label{fig:saprg}
Plots of the estimates for the amplitude combination $8BF$.
}
\end{figure}

Next we study the asymptotic form of the coefficients $r_n=8n^2 p_n \ave{R^2}_n$ 
for the radius of gyration. The generating function  has 
critical exponent $-(\alpha+2\nu)=-2$, so the leading correction-to-scaling term
no longer becomes part of the analytic background term.
We thus use the following asymptotic form: 

\begin{equation}\label{eq:rgsasymp}
r_n \sim \mu^n n \left [ 8BF + \sum_{i\geq 0} a_i/n^{1+i/2} \right ] +
(-1)^{n/2}\mu_-^n n \sum_{i=0}^k b_i/n^i.
\end{equation}
\noindent
In figure~\ref{fig:saprg} we plot the resulting estimates for the amplitude $8BF$.
The predicted exact value \cite{CG93} is 
$BF=\frac{5}{32\pi^2}\sigma a_0=\frac{15\sqrt{3}}{64\pi^2}=0.0411312745\ldots$,
where for the honeycomb lattice $\sigma=2$ and $a_0=3\sqrt{3}/4$. 
Clearly extrapolation of these numerical results yield estimates
consistent with the theoretical prediction.

\subsection{Amplitude ratios of area-weighted moments}

\begin{figure}
\begin{center}
\includegraphics[scale=0.65]{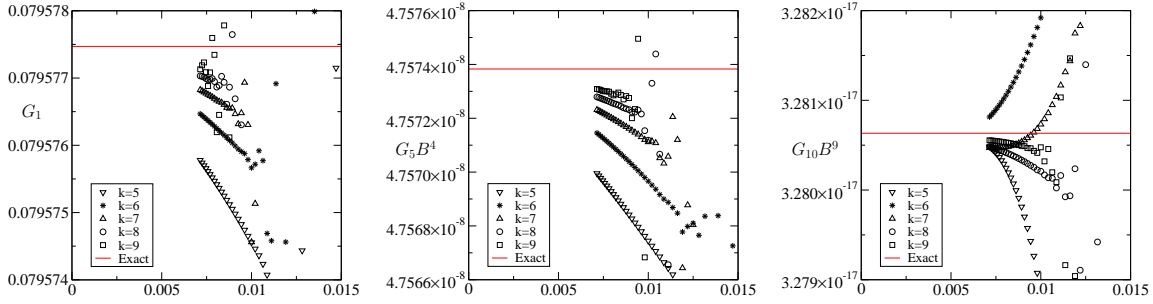}
\end{center}
\caption{\label{fig:momampl}
Plots of the estimates for some of the amplitude combinations $G_m B^{m-1}$.
}
\end{figure}

The amplitudes of the area-weighted moments were studied in \cite{RJG03}.
We fitted the coefficients to the assumed form
\begin{equation}\label{eq:momampl}
n p_n \ave{a^m}_n \sim \mu^n n^{2m\nu+\alpha-2} m! 
\left[ G_m+\sum_{i\ge 0}^k a_i/n^{1+i/2} \right] +
(-1)^{n/2}\mu_-^n n^{2m\nu+\alpha-2} \sum_{i=0}^k b_i/n^i,
\end{equation}
where the amplitude $G_m=G^{(m)}B/m!$ is related to the amplitude
defined in equation (\ref{eq:asympmom}).
The scaling function prediction for the amplitudes $G_m$ is \cite{RGJ01}
\begin{equation}
G_{2m}B^{2m-1} = -\frac{c_{2m}}{4\pi^{3m}} \frac{(3m-2)!}{(6m-3)!}, \qquad
G_{2m+1}B^{2m} = \frac{c_{2m+1}}{(3m)!\pi^{3m+1}2^{6m+2}},
\end{equation}
where the numbers $c_m$ are given by the quadratic recursion 
\begin{equation}
c_m + (3m-4) c_{m-1} + \frac{1}{2}\sum_{r=1}^{m-1} c_{m-r}c_r=0, \qquad c_0=1.
\end{equation}
In figure~\ref{fig:momampl} we have plotted the resulting estimates for
some of the amplitude ratios. Clearly the numerical results are fully
consistent with the theoretical predictions.

\section{Summary and conclusions \label{sec:sum}}

In this paper we have studied series for self-avoiding walks and polygons 
on the honeycomb lattice, including series for metric properties and moments of the 
area-distribution for polygons. We used various methods from Tony Guttmann's tool-kit
to analyse the series. The connective constant, critical exponents and many 
amplitude combinations are known exactly, making it the perfect test-bed for series 
analysis techniques. 

In section~\ref{sec:DA} we used differential approximants to obtain estimates
for the singularities and exponents of the SAP and SAW generating functions. 
Analysis of the SAP series (section \ref{sec:DAsap}) yielded very accurate estimates
for the critical point $x_c^2=0.2928932186(5)$ and exponent $2-\alpha=1.5000004(10)$. 
The estimates agree with the conjectured exact values 
$x_c^2=1/(2+\sqrt{2})= 0.292893218813\ldots$ and $2-\alpha=3/2$. In addition
we found clear evidence of a non-physical singularity on the negative
axis at $x=x_-=-0.412305(5)$ with an associated critical exponent $\alpha_-=3/2$.
The analysis of the SAW series
(section \ref{sec:DAsaw}) also yielded estimates consistent with the predictions
of the exact values. In this case there was a non-physical singularity at
$x=-x_c$ as well as a pair of complex singularities at $x=\pm 0.64215(15) {\rm i}$. 
So the excellent agreement is particularly impressive in light of the 
quite complicated functional form of the generating function which has at least
three non-physical singularities as well as non-analytical corrections to scaling.
So the walk series is obviously much harder to analyse and it is 
a great testament to the method of differential approximants that the analysis
nevertheless yields such accurate estimates.

In section~\ref{sec:Ampl} we looked closer at  the asymptotic form of the
coefficients. In particular we obtained accurate estimates for the leading 
critical amplitudes  $A$ and $B$. Our method of analysis consisted in fitting 
the coefficients to an assumed asymptotic form. In section~\ref{sec:Bampl} we
analysed the SAP  series and demonstrated clearly that in fitting to the
coefficients we cannot ignore the singularity at $x=x_-$ even though it is
exponentially suppressed asymptotically. After inclusion of this term estimates
for the leading amplitude $B$ were well behaved when including only analytic
corrections and we found $B=1.2719299(1)$. We argued that this behaviour was 
consistent with a corrections-to-scaling exponent $\Delta_1$ being half-integer 
valued and in particular consistent with the prediction by Nienhuis that
 $\Delta_1=3/2$ (in section~\ref{sec:corr} we showed the absence of a term with 
$\Delta_1=11/16$). In the analysis of the SAW series we discovered some subtleties 
regarding the inclusion of the terms arising from the complex pair of singularities. 
Despite a quite complicated asymptotic form (\ref{eq:Afinal}) taking into account
all the singularities and the corrections-to-scaling exponent $\Delta_1=3/2$ we 
could still obtain a quite accurate amplitude estimate  $A=1.1449355(5)$. This analysis
clearly shows that it is possible to probe quite deeply into the asymptotic
behaviour of the series coefficients and in particular to distinguish between
different corrections to scaling.

\section*{E-mail or WWW retrieval of series}

The series for the generating functions studied in this paper 
can be obtained via e-mail by sending a request to 
I.Jensen@ms.unimelb.edu.au or via the world wide web on the URL
http://www.ms.unimelb.edu.au/\~{ }iwan/ by following the instructions.

\section*{Acknowledgments}

The calculations in this paper would not have been possible
without a generous grant of computer time from the
Australian Partnership for Advanced Computing (APAC). We also used
the computational resources of the Victorian Partnership for Advanced 
Computing (VPAC). We gratefully acknowledge financial support from 
the Australian Research Council.

\end{document}